\documentclass[preprint,prb,showpacs]{revtex4}

\usepackage{graphicx}
\usepackage{amsfonts}
\usepackage{amssymb}
\usepackage{amsmath}

\usepackage{color}

\usepackage{epsfig}

\newcommand{\beq}{\begin{equation}}
\newcommand{\eeq}{\end{equation}}
\newcommand{\avg}[1]{\left< #1 \right>}

\newcommand{\barr}{\begin{eqnarray}}
\newcommand{\earr}{\end{eqnarray}}
\newcommand{\Ord}[1]{{\cal O}\left( #1\right)}

\def\floor#1{\lfloor #1 \rfloor}

\def\set#1{\{ #1 \}}

\begin{document}

\title{Cavity method for quantum spin glasses on the Bethe lattice}

\author{C. Laumann}

\affiliation{\emph{Department of Physics}, \emph{Joseph Henry
    Laboratories},\emph{Princeton University}, Princeton NJ 08544}

\affiliation{\emph{Princeton Center for Theoretical Physics, Princeton
    University}, Princeton NJ 08544}

\author{A. Scardicchio}

\affiliation{\emph{Department of Physics}, \emph{Joseph Henry
    Laboratories},\emph{Princeton University}, Princeton NJ 08544}

\affiliation{\emph{Princeton Center for Theoretical Physics, Princeton
    University}, Princeton NJ 08544}

\author{S. L. Sondhi}

\affiliation{\emph{Department of Physics}, \emph{Joseph Henry
    Laboratories},\emph{Princeton University}, Princeton NJ 08544}

\affiliation{\emph{Princeton Center for Theoretical Physics, Princeton
    University}, Princeton NJ 08544}

\begin{abstract}
  We propose a generalization of the cavity method to quantum spin
  glasses on fixed connectivity lattices. Our work is motivated by the
  recent refinements of the classical technique and its potential
  application to quantum computational problems. We numerically solve
  for the phase structure of a connectivity $q=3$ transverse field
  Ising model on a Bethe lattice with $\pm J$ couplings, and
  investigate the distribution of various classical and quantum
  observables.
\end{abstract}

\pacs{75.50.Lk, 05.30.-d}

\maketitle

\section{Introduction}

The appearance of finite connectivity trees (\emph{eg} Cayley trees,
Bethe lattices) in the study of spin glasses has a long history that
goes back to the first papers on the Sherrington-Kirkpatrick model
\cite{SK,TAP}. Thouless, Anderson and Palmer\cite{TAP} showed by means
of a diagrammatic expansion of the partition function that in the
spin-glass phase the mean-field theory is defined by the diagrams
which describe an infinite tree with connectivity $q\gg 1$. They then
wrote the mean field equations%
\footnote{It is interesting to notice that in the TAP
  approach\cite{TAP}, the free energy was found as a consequence of
  the mean field equations, reversing the usual order of
  reasoning. This was mended by Anderson shortly
  thereafter\cite{Anderson} and in many other papers later. However
  this unusual, secondary role of the free energy in the problem at
  hand has survived until today and seems to be a peculiarity of spin
  glasses.}%
for such a tree and simplified the results in the small coupling,
large connectivity regime (since in the SK model $q=N$ and the
couplings $J_{ij}\propto 1/\sqrt{N}$). Those mean field equations are
known as the TAP equations and the peculiar characteristics of their
solutions, in particular their large number \cite{BrayMoore}, were an
important indicator of the complexity of the spin glass phase that
arose in parallel with the remarkable developments in the study of the
replicated free energy \cite{AT,ParisiRSB}.

Many authors have since studied the problem of a spin glass on the
Bethe lattice \cite{VB,Bowman,Kwon,Gold1,chayes} with two distinct
motivations. The first of these has been to attempt to find a model of
a short ranged spin glass where one can rigorously assess Parisi's
picture for the organization of the Gibbs states in short-range
systems \cite{Huse}.  The second has come from computer science
whereby a set of optimization problems can be recast as frustrated
problems on random graphs with the local connectivity of a tree
\cite{Ksat}.

Despite much work, the Bethe lattice has not yielded a decisive
verdict for or against the Parisi picture of a multitude of Gibbs
states in the ordered phase of a spin glass although the case for it
is perhaps stronger here than on regular lattices with short ranged
interactions. This is tied up with the question of defining the
infinite Bethe lattice limit starting from finite graphs. It is
possible to do so either via a sequence of Cayley trees with random,
frustrating, boundary conditions or via a sequence of random graphs
with fixed connectivity. The latter sequence has frustrating loops of
typical size diverging $O(\log(N)/\log(q-1))$ for a graph with $N$
points and connectivity $q$ and thus locally looks like a tree
\cite{Bollobas}. The Cayley tree sequence can always be analyzed in
terms of a recursion relation that we review below and does not appear
to lead to the Parisi structure \cite{chayes}. However, it has been
argued that the random graph problem is fundamentally different and
{\it does} lead to replica symmetry breaking \cite{MP}.

Happily, the perspective provided by spin glass theory on optimization
problems has been quite fruitful regardless. Starting in the early
80's \cite{Kirk,FuAnderson}, it became clear that much was gained by
the recognition that various optimization problems in computer science
were equivalent to finding ground states of certain statistical
mechanics problems. A typical example of this connection is given by
the k-SAT problem, which asks ``Given a boolean expression $J$ on $N$
bits $\sigma_i$ composed of the conjunction of $M$ clauses, each of
which involves exactly $k$ of the bits, is it satisfiable by some bit
assignment?''  This can be recast in Hamiltonian form by writing a
cost function $H_J[\set{\sigma_i}]$ that evaluates the number of
violated clauses. In this language, the bits naturally become Ising
spins, the expression $J$ becomes a particular instance of some spin
glass and the original question requires determining the ground state
energy. The large $N$ limit is a problem in statistical mechanics and
typical-case analysis for k-SAT becomes the disorder averaged analysis
of spin glass theory. In this limit, $k$-SAT develops several phase
transitions as a function of $\alpha = M/N$, the number of clauses per
bit\cite{KsatNew}. With increasing $\alpha$, the most salient
features are that the problem goes from being easily solved and
satisfiable, to an intermediate glassy phase with many local ground
states, to a typically unsatisfiable phase where the ground state
energy density is positive.

A key role in these developments has been played by the so called
\emph{cavity method}---a complex of analytical and numerical
techniques refined recently \cite{MP, Ksat}---for studying classical
spin glasses on tree-like graphs.  Applied to the k-SAT problem, the
cavity method suggests the above phase diagram and provides numerical
values for its critical points. Moreover, the technique can be applied
to a particular instance of k-SAT and the information so obtained
about the free energy landscape now guides the search procedure in
state-of-the-art k-SAT algorithms \cite{Ksat}.

In this paper we turn to \emph{quantum} spin glasses on Bethe
lattices, specifically to the problem of extending the cavity method
to their analysis.  As in the classical case, there are two distinct
reasons to be interested in these systems. There is the intrinsic
interest of the interplay between quantum mechanics and spin glass
behavior, about which the difficulties of the classical case serve as
both caution and enticement. A second motivation now arises from the
rapid recent developments in quantum computing. In particular, the
discovery of the ground state of a classical spin system $H_J$ derived
from a computational problem is ideally suited to solution by the
\emph{adiabatic algorithm} \cite{AdiabAlg}, which allows a quantum
computer to solve such a problem by means of an adiabatic change of
the parameters in the Hamiltonian.  The adiabatic algorithm would
trace a path in operator space starting from a simpler Hamiltonian
$H_0$ and ending at $H_J$. For instance, $H_0$ could be the
Hamiltonian for $N$ independent spins in a large transverse magnetic
field $B_t$ and the path could be to slowly lower the field and raise
the Ising couplings $J$. Starting from the easily found ground state
of $H_0$ a sufficiently slow evolution will carry the system to the
ground state of $H_J$. The power of this algorithm is therefore
measured in the scaling of the evolution time with the number of
variables $N$.

This discussion certainly suggests that a careful study of the phase
transitions encountered on the path between $H_0$ and $H_J$ is in
order.  However, the necessity of understanding the ``deep'' quantum
spin glass phase far from the phase transition is also clear. The
structure of this phase, especially its nontrivial energy landscape,
is probably as important as the nature of the phase transition. In
this paper we take a first step toward understanding the quantum spin
glass phase by generalizing the cavity method used to study the
classical problem. Compared to the replica method or direct study of
the quantum TAP equations, we believe the cavity method outlined in
this paper provides much more physically transparent information about
the SG phase. It also may be applied to many other quantum phase
transitions on the Bethe lattice, whether induced by disorder or more
typical symmetry breaking.

We note that the topic of quantum spin glasses on trees has been
tackled before in the literature from the point of view of statistical
mechanics \cite{Kopec} and computation theory
\cite{Smelyanskiy}. However, in order to proceed analytically and due
to the complexity of the problem, these works have employed several
uncontrolled approximations. We will compare with them briefly in the
core of the paper. While we have used approximations in our
(numerical) analysis as well, we believe they are better controlled,
since we show how to systematically improve them and how the results
are robust with respect to the improvements. More evolved
computational methods for tackling very similar problems have appeared
in the literature on dynamical mean field theory \cite{DMFT} and could
possibly be applied to the quantum cavity method on glasses. These
methods may allow a study of the zero temperature case, which our
approximations could not capture.

Let us turn now to a very brief overview of the results in this
paper. We first formulate the cavity method for the transverse Ising
spin glass on a Bethe lattice. This is a Markov process for the
on-site effective action whose stationary probability distribution can
be found numerically using a population dynamics algorithm. This will
be the main result of the paper. In this way we obtain the phase
boundary in the $(B_t,T)$ plane, the values of the usual thermodynamic
quantities (free energy, energy, entropy and $q_{EA}$, the
Edward-Anderson order parameter \cite{EA}) and more typical `quantum'
quantities like the single spin von Neumann entropy. Indeed, from the
fixed point probability distributions for the effective actions we can
calculate distributions for all of the statistical properties of the
system. This leaves much room for further work.

The plan of the paper is the following: we introduce the classical
cavity method for uniform ferromagnetic systems and spin glasses in
Section \ref{sec:classcav}; we propose its generalization to the
quantum spin glass problem in \ref{sec:quancav}; we apply it to the
numerical study of a Bethe lattice with connectivity three in
\ref{sec:numerics}. Discussion and directions for further work will be
presented in the last sections.

\section{Classical Cavity Method}
\label{sec:classcav}
 
The cavity method is a way of finding the spin glass free energy by
means of self-consistency equations for the probability distributions
of quantities characterizing the statistics of the spins (the
\emph{cavity fields}). It has several virtues compared with the
replica method, in particular if applied to spin glasses with finite
connectivity. We will return to them after we have explained how the
cavity method works.

\subsection{The Bethe-Peierls method for the ferromagnet}

For conceptual clarity we will first recall the ``cavity method'' in
its simplest form as applied to a ferromagnetic Ising model living on
a Bethe lattice with connectivity $q$. As mentioned above, there is
considerable subtlety in defining a sensible infinite tree model, but
for this simple case, a sequence of Cayley trees with uniform small
boundary fields to break the Ising symmetry will suffice. In any
event, we shall focus here on the local structure of the model.

The classical Hamiltonian is
\begin{equation}
\label{eq:classham}
H=-\sum_{(ij)}J_{ij}\sigma_i\sigma_j
\end{equation}
where $\sigma_i\in\set{\pm{}1} $ are Ising spins, $J_{ij} = J > 0$ is
a uniform ferromagnetic coupling and the sum is over bonds in the
Bethe lattice. To create a \emph{cavity}, we pick a spin $\sigma_0$ in
the Bethe lattice and imagine removing it:
\begin{equation}
\includegraphics{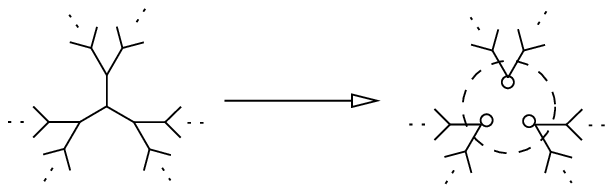}
\end{equation}
Each of $\sigma_0$'s neighbors is a cavity spin, connected to $q-1$
spins and sitting at the root of a branch of the original tree. Notice
that in the absence of $\sigma_0$ these $q$ branches are entirely
independent. Similarly, a cavity spin $\sigma_1$ mediates the only
interaction between the $q-1$ branches sitting above it.

\begin{figure}[h]
\begin{center}
\includegraphics{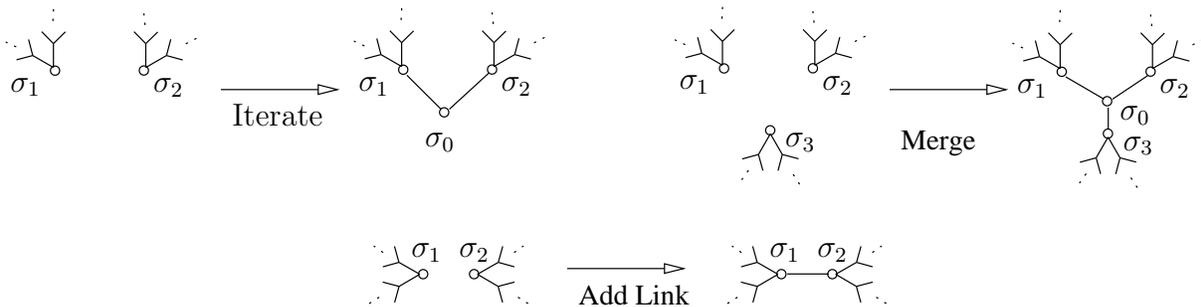}
\caption{The three cavity operations. Although we have not labeled
  them in the Figure, the new links are $J_{01}$, $J_{02}$, etc ...}
\label{fig:cavity-ops}
\end{center}
\end{figure}

We can define three important operations we can perform on graphs with
cavity spins: \emph{iteration}, \emph{merging}, and \emph{link
  addition} (Fig. \ref{fig:cavity-ops}.) For our immediate purposes,
the most important of these is iteration, which takes $q-1$ rooted
branches of depth $l$ and links them into a single new branch of depth
$l+1$ with a new cavity spin at the root. Thus we can construct an
arbitrarily large Cayley tree by iterating inward from its boundary
spins many times followed by a final merge operation to form the
center.

Consider the iteration operation: the added spin $\sigma_0$ receives
thermodynamic information regarding each of the $q-1$ branches only
through the thermal distribution of the cavity spins $\sigma_1, ...,
\sigma_{q-1}$. In the absence of $\sigma_0$, each of these Ising
variables has independent statistics characterized fully by its
thermal probability distribution:
\begin{equation}
\psi_i(\sigma_i)=\frac{e^{\beta h_i \sigma_i}}{2\cosh(\beta h_i)}.
\end{equation}
which defines the \emph{cavity field} $h_i$.  Since $\sigma_1 \in
\set{\pm{}1}$, there are only two possible configurations of the spin
and only $2-1=1$ real numbers are needed to characterize the
probability distribution. In this sense, the cavity field $h_i$ is
merely a good parameterization of the distribution
$\psi_i(\sigma_i)$. We emphasize this viewpoint because it will
naturally generalize to the quantum case.

We introduce a simple graphical convention for cavity spins:
\begin{equation}
\includegraphics{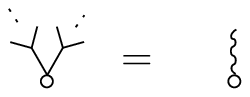}
\end{equation}
The open circle indicates a spin variable and the wiggly line
indicates the effective field attached to it. With this notation, an
iteration operation can be represented:
\begin{equation}
\includegraphics{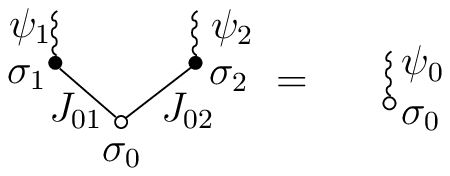}
\end{equation}
where the filled circle indicates summing out a spin variable. More
formally, the state of the spin $\sigma_0$ depends on the state of the
$q-1$ spins as
\begin{equation}
  \label{eq:consistencypsi}
\psi_0(\sigma_	0)=\frac{1}{Z}\sum_{\sigma_1,...,\sigma_{q-1}=\pm 1}\exp\left(\beta \sum_i J_{0i}\sigma_0\sigma_i\right)\psi_1(\sigma_1)...\psi_{q-1}(\sigma_{q-1}),
\end{equation}
where $Z$ is a normalization factor so
$\sum_{\sigma_0}\psi(\sigma_0)=1$. In terms of cavity fields this
equation implies
\begin{equation}
\label{eq:rech}
h_0=\frac{1}{\beta}\sum_{i=1}^{q-1}\tanh^{-1}(\tanh(\beta J_{0i})\tanh(\beta h_i))\equiv U(\{h_i\},\{J_{0i}\}).
\end{equation}

Finally, to solve the ferromagnet, we note that all the $J_{ij} = J$
and apply uniform boundary fields (say slightly positive) to the
Cayley tree ``leaves''. From this uniform starting point, we expect to
find fixed points for the cavity fields under iteration given by
\begin{equation}
\label{eq:bethepeierls}
h = \frac{q-1}{\beta} \tanh^{-1}(\tanh(\beta J) \tanh(\beta h)) 
\end{equation}
This is precisely the Bethe-Peierls self-consistency equation for a
mean field ferromagnetic in a lattice of coordination number $q$. 

\subsection{Classical Spin Glass on a Cayley Tree}

The careful reader will have noticed that until Eq.
(\ref{eq:bethepeierls}), we did not anywhere exploit the uniformity of
$J_{ij}$ or $h_i$ in the foregoing discussion. With this foundation
laid, we can make short work of the classical Bethe lattice Ising spin
glass. Again, we consider a Bethe lattice defined as a limit of Cayley
trees with fixed boundary conditions rather than as a limit of random
graphs. The Hamiltonian is now
\begin{equation}
H_J = -\sum_{(ij)} J_{ij} \sigma_i \sigma_j
\end{equation}
where the $J_{ij}$ are i.i.d. random variables drawn from some
distribution $P(J)$. For simplicity and because of its connection to
computational problems, we will restrict our attention to the $\pm J$
model
\begin{equation}
\label{eq:plusminusJ}
P(J_{ij})=\frac{1}{2}\delta(J_{ij}-J)+\frac{1}{2}\delta(J_{ij}+J),
\end{equation}
although much of the discussion has broader validity.

The iteration equation (\ref{eq:consistencypsi}) is still valid for
particular realizations of the $J$, but since these are random
variables, it now defines a Markov process for the cavity
fields. Throughout the graph, these fields will be site dependent
random variables, but \emph{deep inside the tree}, they ought to be
distributed according to a probability distribution $P(h)$ that
represents a fixed point of the Markov process.%
\footnote{In particular, the cavity field distribution will depend
  only on the depth of a site in the Cayley tree, assuming
  i.i.d. boundary fields. This distribution becomes depth independent
  sufficiently deep in the tree.}  This fixed point distribution will
satisfy
\begin{equation}
\label{eq:consP}
P(h)=\int \prod_{i=1}^{q-1} dh_i P(h_i)\avg{\delta(h-U(\{h_i\},\{J_{0i}\})}_J.
\end{equation}
In terms of the spin distribution, this becomes the functional equation
\begin{equation}
P[\psi] = \int \left(\prod_{i=1}^{q-1} D\psi_i P[\psi_i] \right) %
 \avg{ \delta[ \psi(\sigma) - \psi_0(\sigma; \set{\psi_i},\set{J_{0i}})]}_J
\end{equation}
where $\psi_0(\sigma)$ is given by Eq. (\ref{eq:consistencypsi}).

This distribution is the order parameter for the spin glass. It is a
$\delta$ function at $h = 0$ in the high temperature phase. As the
temperature is lowered, $P(h)$ broadens to have finite support below
some mean field-like phase transition. Defining $\tau=T/J$, this phase transition is located at $\tau_c =1/
\tanh^{-1}(\frac{1}{\sqrt{q - 1}})$. It is possible to write the free
energy per site in terms of $P(h)$ as
\begin{eqnarray}
F &=&\int\prod_{i=1}^{q} dh_i P(h_i) F_{q+1}-\frac{q}{2}\int\prod_{i=1}^{2} dh_i P(h_i) F_{2},\\
F_{q+1}&=&-\frac{1}{\beta}\avg{\ln\sum_{\sigma_0,\sigma_1,...,\sigma_{q}}\exp(\beta\sum_{i=1}^qJ_{0i}\sigma_0\sigma_i)\prod_{i=1}^q\psi_i(\sigma_i)}_{J,\psi},\\
F_2&=&-\frac{1}{\beta}\avg{\ln\sum_{\sigma_1,\sigma_2}\exp(\beta{}J_{12}\sigma_1\sigma_2)\psi_1(\sigma_1)\psi_2(\sigma_2)}_{J,\psi}.
\end{eqnarray}
This rather complicated looking expression is actually simply the
average change in free energy due to a merge operation minus
$\frac{q}{2}$ times the average change in free energy due to a link
addition. Graphically,
\begin{equation}
\label{eq:free-energy}
\includegraphics{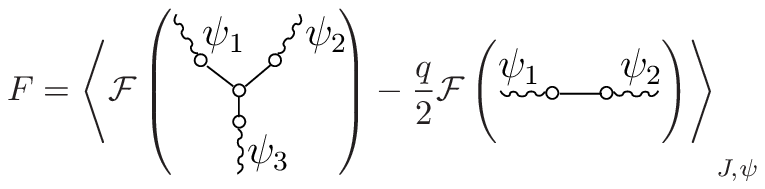}
\end{equation}
where ${\mathcal F}$ of a diagram is the free energy of a system with
spin variables given by the unfilled circles.

It is possible to see that if $P$ satisfies (\ref{eq:consP}) then
$\delta F/\delta P(h)=0$. Other expressions for the free energy $F$
have appeared in the literature but it has been shown that they are
all equivalent to each other, if the consistency equation
(\ref{eq:consP}) is verified. The role of the free energy in the
cavity method is secondary as one does not solve the variational
problem, as in the replica method, by working on the free energy
directly. Rather, one finds the probability distribution $P$ by
analytical or numerical methods and then derives all of the
statistical observables from $P$.%
\footnote{Readers familiar with the literature may note that this
  allows one to bypass the parameterization of the order parameter on
  the replica indices, a particularly thankless task on the Bethe
  lattice since an infinite sequence of order parameters $q_{a,b},\
  q_{a,b,c},\ q_{a,b,c,d},...$ is necessary \cite{Mottishaw}.} The
equivalence of the two formulations has been put forward in Ref.
\onlinecite{MP,Gold1} and in many other works.

We have up to this point assumed that the underlying lattice is in
fact a tree and that the removal of a spin to create a cavity
completely decouples the neighboring branches. On such models, the
cavity method we have described is exact. An important generalization
of the cavity method arises in its heuristic application to locally
tree-\emph{like} random graphs where the typical loop size diverges
logarithmically and any finite neighborhood is a simple tree. In this
case however, the decoupling of cavity spins is not exact and it is
necessary to introduce the so-called ``replica symmetry breaking''
ansatz on the structure of thermodynamic states in order to correctly
describe the frozen (glassy) phase. Although this is a vital component
of the modern understanding of such glasses, we have not included it
in our quantum treatment below because we believe their is still much
to be understood about the quantum model on the much simpler trees. We
refer the interested reader to Ref.~\onlinecite{MP} for a more
detailed description of replica symmetry breaking in the classical
cavity method.

Now, having introduced the cavity method in classical statistical
mechanics we go on to generalize it to quantum mechanics.

\section{Quantum Cavity Method}
\label{sec:quancav}

\subsection{Exact Framework}

We consider the modification of the Hamiltonian (\ref{eq:classham})
due to the introduction of a transverse magnetic field
\begin{equation}
\label{eq:quantum-hamiltonian}
H=-\sum_{(ij)}J_{ij}\sigma_i^z\sigma_j^z - B_t \sum_i \sigma_i^x.
\end{equation}
This is called the \emph{transverse field Ising spin glass} in the
literature. The Ising variables of the previous section have been
replaced by Pauli matrices $\sigma^z$ and the magnetic field couples
to the matrices $\sigma^x$. The fact that $\sigma^z$ and $\sigma^x$ do
not commute gives rise to a host of interesting new features due to
the interplay of quantum mechanics and disorder \cite{Fisher}.

The usual Suzuki-Trotter decomposition allows us to rewrite the
problem in terms of $N_t$ Ising spins per quantum spin, where the
number $N_t$ needs to be sent to infinity eventually. The additional
dimension which is introduced in this way is the usual imaginary
time. The $\sigma_i^z\sigma_j^z$ interactions are time-translation
invariant (the disorder is correlated in the time direction) while the
$\sigma^x$ terms give a ferromagnetic nearest-neighbor interaction in
the time direction. Before writing the Hamiltonian let us introduce
some notation.

For any finite $N_t$ we will refer to the Ising spin configuration at
a given site $i$ as a ``rod'' of spins. The rod at site $i$ is
described by $N_t$ spins $\sigma_i(t)$ where $t$ takes values from $0$
to $\beta$ in steps of $\Delta t=\beta/N_t$, with periodic boundary
conditions $\sigma(0)=\sigma(\beta)$. This notation is convenient if
the limit $N_t\to\infty$ is eventually performed, since the rod is
represented by a function $\sigma(t):[0,\beta]\to\{-1,1\}$ with
$\sigma_i(0)=\sigma_i(\beta)$. The rod statistics are described by a
probability distribution $\psi[\sigma(t)]$, a functional of
$\sigma(t)$, that gives a positive real number for every configuration
$\sigma(t)$. The normalization condition reads $\sum_{\{\sigma(t)\}
}\psi[\sigma(t)]=1$.

The partition function is written as
\begin{equation}
\label{eq:partfunQ}
Z=\sum_{\{\sigma_i(t)\}} e^{-\beta H[\sigma]}
\end{equation}
where the Hamiltonian is:
\begin{equation}
\label{eq:hamiltQ}
\beta H=-\sum_t\sum_{(ij)}\Delta t J_{ij}\sigma_i(t)\sigma_j(t)-\Gamma\sum_t \sum_i \sigma_i(t)\sigma_i(t+\Delta t).
\end{equation}
We can also write this as a sum over links of the energy of a link:
\begin{equation}
\label{eq:link-hamiltonian}
\beta H_{ij}=-\sum_t\Delta t J_{ij}\sigma_i(t)\sigma_j(t)-\frac{1}{q}\Gamma\sum_t \sigma_i(t)\sigma_i(t+\Delta t)
\end{equation}
where a fraction $1/q$ of the imaginary-time interaction is associated
to each link (there are $q$ links per spin). Here
$\Gamma=\frac{1}{2}\ln\coth(\beta B_t/N_t)$.

Notice that $\Gamma>0$ so the system is ferromagnetic, and moreover
when $\beta B_t/N_t\ll 1$ then $\Gamma\gg 1$ and the coupling along
the time direction is strongly ferromagnetic. In particular, for
$B_t=0$ the spins in any given rod are locked together as $N_t$
useless copies of a single Ising spin.  Thus, the results reduce to
the classical case smoothly.

\begin{figure}[h]
\begin{center}
\includegraphics{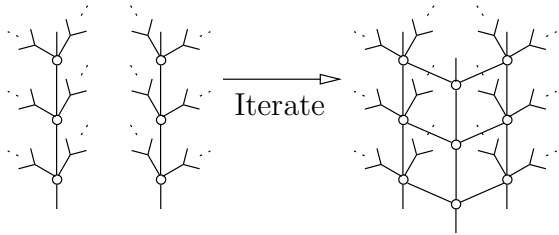}
\caption{Iteration of cavity rods. There are periodic boundary
  conditions in the imaginary time (vertical) direction. Other cavity
  operations are analogous.}
\label{fig:tree-iteration}
\end{center}
\end{figure}

The spatial tree-like structure of the original problem is reflected
in the tree-like structure of the interaction between rods. We can
therefore imagine an iteration process with rods replacing the spins,
in which we have $q-1$ \emph{cavity rods} $\psi_i[\sigma_i(t)]$ which
are merged and determine the state of the rod $\psi_0 [\sigma_0(t)]$
(see Fig. \ref{fig:tree-iteration}). This corresponds to a recursion
relation for the calculation of the partition function of the
branches, analogous to the classical equation
(\ref{eq:consistencypsi}):
\begin{equation}
\label{eq:recrod}
\psi_0[\sigma_0(t)]=\frac{1}{Z}\sum_{\{\sigma_{i\geq 1}(t)\} }e^{-\sum_{i=1}^{q-1}\beta H_{0i}}\prod_{i=1}^{q-1}\psi_i[\sigma_i(t)].
\end{equation}

Just as in the classical case, this iteration equation is already
enough to solve an interesting ferromagnetic problem. Consider the
case $q=2$ and $J_{ij}=J$. For $q=2$ our Bethe lattice is a simple
chain of spins, moreover all $J$'s being equal we can take $J>0$
without loss of generality. We then recover the well known
ferromagnetic Ising \emph{chain} with a transverse field, an exactly
solvable system. One way to solve it is to use Onsager's transfer
matrix method. In fact, the iteration equation (\ref{eq:recrod}) can
be rewritten as just such a transfer matrix equation where $\psi$ is
the $2^N$ dimensional vector and $e^{-\beta H_{01}}$ the transfer
matrix $T$:
\begin{equation}
Z\psi_0=T\cdot\psi_1.
\end{equation}
The fixed point of this iteration gives the eigenvector $\psi$
corresponding to the largest eigenvalue $Z$. In the limit
$N_t\to\infty$, this is an exact solution which contains all the
information about classical and quantum phase transitions.

Returning to the case of the spin glass, we can write down the quantum
cavity fixed point equation analogous to the classical equation
(\ref{eq:consP}) immediately:
\begin{equation}
\label{eq:rsfixedpoint}
\begin{array}{rcl}
P_{FP}\left[ \psi[\sigma(t)] \right] & = & \avg{\delta\left[ \psi[\sigma(t)] - \psi_0[\sigma(t); \{J_{0i}, \psi_i \}_{i=1}^{q-1}] \right] }_{J_{0i}, \psi_i} \vspace{1em} \\
 & = & \int \left( \prod_{i=1}^{q-1} D\psi_i P_{FP}\left[ \psi_i \right] dJ_{0i} P(J_{0i}) \right) 
 		\delta\left[ \psi[\sigma(t)] - \psi_0[\sigma(t); \{J_{0i}, \psi_i \}_{i=1}^{q-1}] \right]
\end{array}
\end{equation}
where the iterated rod action $\psi_0[\sigma(t); \{J_{0i}, \psi_i
\}_{i=1}^{q-1}]$ is given by Eq. (\ref{eq:recrod}) and $P(J_{0i})$ is the
fixed prior distribution for couplings Eq. (\ref{eq:plusminusJ}). This is
a functional equation for $P_{FP}\left[ \psi[\sigma(t)] \right]$, the
fixed point probability distribution of the effective distribution
describing iterated cavity rods. In the limit $N_t\to\infty$, it is
exact but difficult to solve in closed form. It is certainly possible
that analytic progress can be made, but we have not succeeded thus
far. However, it is amenable to numerical study at finite $N_t$ under
certain approximations and also perhaps by continuous time Monte Carlo
for $N_t\to\infty$. In the remainder of this section we will explore
the finite $N_t$ approach.

We must first parameterize our generic vector $\psi$ in the
$2^{N_t}$-dimensional space of the configurations of the rods. In
principle it is described by $2^{N_t}-1$ real numbers, which can be
reduced by a factor $O(N_t)$ by exploiting time-translation symmetry
and the periodic boundary conditions. A natural way to parameterize it
is in term of the \emph{effective action} of the rod
\begin{equation}
\psi[\sigma(t)]=e^{-S[\sigma]}
\end{equation}
where we expand $S$ in a series of increasing clusters of interacting spins.
\begin{eqnarray}
\label{eq:expanS}
S[\sigma]=&-&\log Z-h \Delta t \sum_t \sigma(t)-\sum_{t,t'}\Delta t^2 C^{(2)}(t'-t)\sigma(t)\sigma(t')-\nonumber\\ &-&\sum_{t,t',t''}\Delta t^3 C^{(3)}(t'-t,t''-t')\sigma(t)\sigma(t')\sigma(t'')+...
\end{eqnarray} 
In principle, the sum includes up to $N_t$-spin interaction terms (the
normalization factor has been included as a spin-independent term in
the effective action). In practice, we truncate the action expansion
at second order to keep the numerical requirements manageable. We 
comment below on the limits in which this truncation is exact. Notice
that $C^{(2)}(t)$ is the kernel for $2$-point in time interactions,
\emph{not} the dynamical two-point correlation function, often denoted
$c^{(2)}(t) = \avg{\sigma(t)\sigma(0)}$.

The functions $h, C^{(i)}$ are random quantities characterized by the
Markov process defined by the \emph{iteration procedure}. By writing
the representations of the vectors $\psi$ in terms of the effective
action (\ref{eq:expanS}) we can rewrite the iteration equation as
\begin{equation}
e^{-S[\sigma,\{h_0,C_0\}]}=\sum_{\{\sigma_1(t)\},...,\{\sigma_{q-1}(t)\}}\!\!\!\!\!\!e^{-\sum_{i=1}^{q-1}\beta H_{0i}}\prod_{j=1}^{q-1}e^{-S[\sigma,\{h_j,C_j\}]}.
\end{equation}
This gives an implicit update map from the `old' $q-1$ parameters
$h_j,C^{(2)}_j,C^{(3)}_j,...$ and the couplings $J_{0j}, B_t$ to the
`new' parameters $h_0, C^{(2)}_0,C^{(3)}_0,...$. The statistics
generated by this Markov process and in particular its fixed point
distribution
\begin{equation}
P(h,C^{(2)},C^{(3)},...)
\end{equation}
are the solution of the problem. 
 
\subsection{Approximations}

\subsubsection{Action Representation}
\label{sec:acti-repr}

By parameterizing the quantum dynamics through the action expansion
(\ref{eq:expanS}) we do not make any \emph{a priori} assumptions about
the nature of the spin-spin correlations in the time direction -- that
is, if we could keep all of the terms in the expansion, it would be an
exact treatment. We avoid in this way the spherical approximation
which has been used in Ref.~\onlinecite{Kopec}, since the results thus
obtained do not reduce to the well-known classical results for
$B_t=0$.

In practice, however, we truncate the cavity actions to second
order. Most usefully, this corresponds to the leading order term in a
large connectivity expansion of the effective action. Indeed, in a
large $q$ treatment, in which the couplings $J_{ij}$ must be scaled as
$1/\sqrt{q}$ for the disordered model, one finds that the one- and
two-body terms in the effective action for the rod at the root of a
tree are $\Ord{1/q^0}$, the three- and four-body terms are
$\Ord{1/q^1}$ and so on\cite{DMFT}. The truncation to second order is
thus both exact and necessary in the $q\to\infty$ limit. We note that
the oft-used \emph{static approximation}\cite{static} is thus
incorrect even at infinite connectivity for disordered models. The
truncation to second order is also exact at high temperature,
regardless of the value of $q$.

Numerical investigations of small systems suggest that the higher
order interactions are quantitatively small more generally, even at
$q=3$. This is especially true at small $B_t$, where the strong
ferromagnetism in the time-like nearest neighbor bonds dominates. We
note that there is also some error in the numerical fitting of an iterated
cavity distribution to a truncated action. This problem is reminiscent
of a \emph{maximum entropy} model for the statistics of signals and
the \emph{inverse Ising problem} of computer science
\cite{Hinton}. However, even in the disordered system, the interaction
of the spins along imaginary time is always ferromagnetic and the
problem does not present the difficulties that usually accompany the
inverse problems in general statistical mechanics. In other words,
although the original problem is fundamentally frustrated, none of
this frustration appears in the single spin dynamics. The frustration
is taken care of in the treatment of the relevant parameters $h,
C^{(i)}$ as random numbers.

For comparison, we note that our approach is closely related to
self-consistent dynamical mean field theory (DMFT) methods, which also
truncate to non-local two point interactions in the single-spin
effective action (these are the so-called Weiss functions or bare
Green's functions of DMFT treatments). However, DMFT techniques cannot be used
straightforwardly for disordered systems. One cannot assume that the
imaginary-time propagator (or action) for a spin is equal to that of
the spins which surround it but rather that they come from the same
probability distribution. If this observation is taken into account
then we expect to recover functional equations for the distribution of
single site Green's functions analogous to those we have proposed for
the single site action.

\subsubsection{Finite Discretization Error}
\label{sec:discretization-error}

Finally, we comment on where in the $(\tau, B_t)$ phase diagram the
fixed $N_t$ approach will give reliable results. As mentioned above,
on the classical line $B_t = 0$, all spins in a rod are locked
together by divergent nearest neighbor interactions and the Trotter
decomposed system (for \emph{any} $N_t$) reduces to the classical
system exactly. In the opposite limit of large $B_t$ at fixed $N_t$,
the planes of the Trotter system decouple as $\Gamma \rightarrow
0$. Each of these planes is an exact copy of the original $B_t = 0$
spin glass model except with couplings $J_{ij} / N_t$. These will
therefore undergo independent thermal phase transitions at $\tau =
\tau_c / N_t$ and no $\tau = 0$ critical field will be detected. This
explains the phenomenon of ``asymptotic critical lines'' that we
note in our finite $N_t$ phase diagrams.

More generally, for the finite $N_t$ approximation to be valid, $N_t >
k_{\textrm{typ}}$ where $k_{\textrm{typ}}$ is the typical number of
kinks in a rod. For a single spin in a transverse field, a
straightforward calculation shows $\avg{k} = 2 \beta B_t \tanh{\beta
  B_t}$, which for large $\beta B_t$ reduces to $\avg{k} \approx 2
\beta B_t$. Within the paramagnetic phase this calculation remains
nearly exact, although in the spin glass phase $\avg{k}$ will be
suppressed by the presence of longitudinal cavity fields. However,
this inequality $N_t > 2 \beta B_t$ remains a good indicator of the
quality of the approximation and agrees with the regions where it
clearly breaks down.

Thus, this expansion is particularly useful close to line $B_t=0$, and
we will see that it gives reasonable results and insights into the
structure of the problem also deep in the quantum spin glass phase. On
the other hand, the other region of interest $T=0, B_t\simeq
B_t^{\textrm{crit}}$ should not be addressed with this expansion,
unless the description of the spins in terms of continuous time functions
$\sigma(t)$ turns out to be treatable in the future. In this paper, we
will not be able to make definitive statements about the nature of the
quantum phase transition which occurs at this point but we are
definitely able to make statements about the nature of the \emph{spin
  glass phase} when quantum effects are not negligible.

In the next section we show how a simple minded Trotter discretization
with $N_t$ relatively small ($N_t=6$ to $11$) delivers a great deal of
information about the quantum spin glass phase.

\section{Numerical Results}
\label{sec:numerics}

We solve the fixed point equation (\ref{eq:rsfixedpoint}) numerically
using a population dynamics algorithm analogous to that of Mezard and
Parisi\cite{MP}.  We represent $P[\psi]$ by a finite population of
$N_{rods}$ rod actions $\{\psi_i[\sigma(t)] = e^{-S_i[\sigma(t)]}\}$,
where the expansion (\ref{eq:expanS}) of $S_i$ is truncated to second
order.  Each rod action is therefore specified by $1 +
\floor{\frac{N_t}{2}}$ distinct numbers $(h, C^{(2)}(t))$ after
exploiting the periodicity of imaginary time.

This population is initialized from an appropriate uniform
distribution and then iterated as follows:
\begin{enumerate}
\item Select $q-1$ rods $\psi_i$ randomly from the population and
  $q-1$ random $J_{0i}$.
\item Use (\ref{eq:recrod}) to calculate the effective action on an
  iterated cavity spin $\psi_0$ from these rods. In principle, higher
  order interactions may be generated but we truncate them by finding
  the second order action that exactly reproduces the free energies of
  a series of domain wall configurations of varying width. 
\item Randomly replace one element of the population with $\psi_0$.
\item Repeat until convergence in some measure of the population, for
  example the order parameter $q_{EA}$.
\end{enumerate}
In practice, this procedure converges quickly deep in either the
glassy or paramagnetic phase but slows near the phase transition as
the flow of $P[\psi]$ under the iteration equation near the
paramagnetic fixed point becomes marginal.

\begin{figure}[tbp]
\begin{center}
\includegraphics{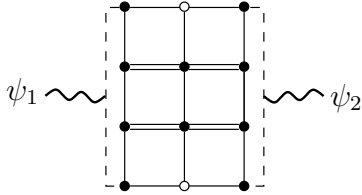}
\caption{Merging of two cavity rods onto a ``broken'' central
  rod. Here $N_t = 3$ but instead of imposing periodic boundary
  conditions everywhere we depict $\sigma(0)$ and $\sigma(N_t)$ as
  independent spins. Each vertical link corresponds to a $\Gamma$
  coupling while each horizontal link corresponds to a $J_{ij}/2N_t$
  coupling. The dashed lines connecting the cavity rods indicate
  identification of the top and bottom spins for each rod and the
  wiggly line indicates an effective action. By imposing different
  values on the top and bottom spin of the broken rod and summing out
  the rest, the elements of the reduced density matrix $\rho_0$ may be
  determined. }
\label{fig:merge-broken}
\end{center}
\end{figure}

Given $P_{FP}[\psi]$, we can calculate the sample averaged free energy
density and local observables such as the link energy, site
magnetization, Edwards-Anderson order parameter $q_{EA} =
\avg{\avg{\sigma_i}_{T}^2}_{i}$ and reduced von Neumann entropy
$S_{vonN}=-\textrm{tr}\rho_0 \log_2(\rho_0)$ by standard Monte Carlo
sampling of these quantities. The free energy density is given by
equation (\ref{eq:free-energy}). The reduced entropy and transverse
magnetization may be derived in the usual way from the reduced density
matrix $\rho_0$ for a spin. We calculate this by performing a merge
operation onto a ``broken rod'' (see Fig. \ref{fig:merge-broken}),
in which periodic boundary conditions are not enforced. The various
elements of the reduced density matrix correspond to imposing
different values on the top and bottom spin of the broken rod and
summing out all other spins in the partition function. Finally,
calculating the average internal energy can be done by averaging the
Hamiltonian (\ref{eq:quantum-hamiltonian}).

\subsection{Numerical Results}

\subsubsection{Phase Diagram}

\begin{figure}[htbp]
\begin{center}
\epsfig{file=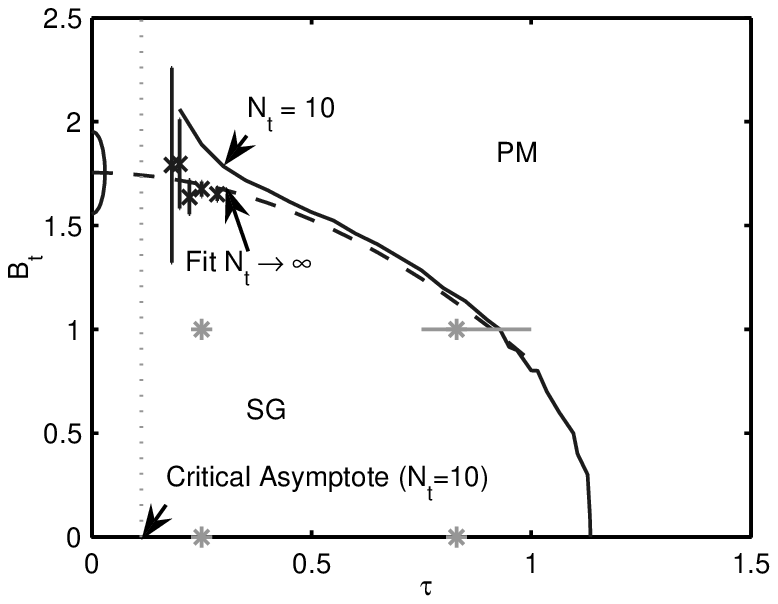,scale=.9}
\epsfig{file=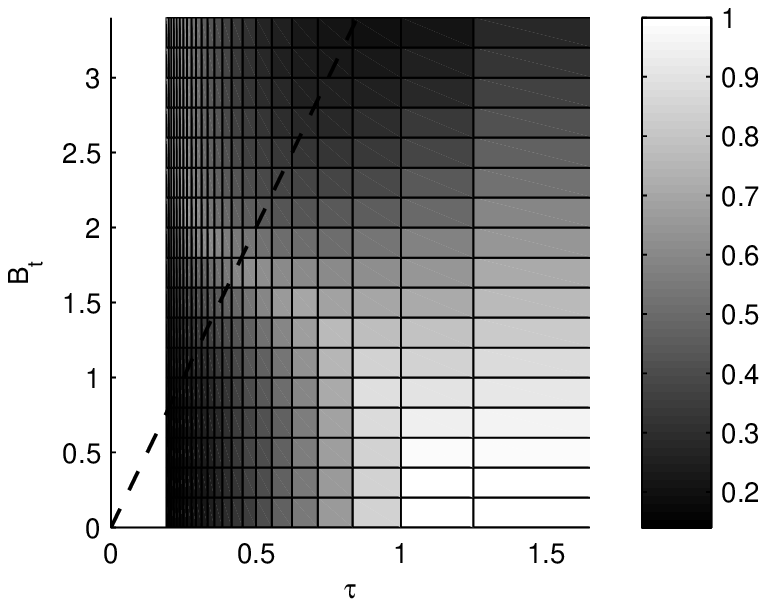,scale=.9}
\caption{(a) Phase diagram at $q=3$. The solid phase transition curve
  has been calculated at $N_t = 10,\ N_{rods} = 2500,\ N_{iter} = 1000
  N_{rods}$ on a fine mesh in the $(\tau, B_t)$ plane. The vertical
  dotted line is the asymptotic critical line for large $B_t$ at $N_t
  = 10$ (\emph{ie} $\tau = \tau_c / N_t$). The points marked x with
  error bars indicate $N_t\rightarrow\infty$ fits based on Fig.
  \ref{fig:trans-curve}. The dashed transition curve is a weighted
  quadratic fit through the estimated low temperature points and the
  $N_t=10$ points in the range $0.5 < \tau < 1$. This leads to an
  estimated $B_t^c = 1.775\pm0.03$. As this fit is clearly heuristic,
  we have suggested a much larger range for our estimate of $B_t^c$ in
  the Figure. The stars and stripes indicate points in the phase space
  which we have investigated in more detail below. (b) The average von
  Neumann entropy $S_{vonN}$ (in bits) of a central spin as a function
  of $(\tau, B_t)$ at $N_t=8$. The dashed line indicates the estimated
  region of validity of the discretization approximation ($B_t \le N_t
  \tau / 2$).}
\label{fig:phase-diag}
\end{center}
\end{figure}

We present numerical results for an investigation of the $q=3$
connectivity model using a naive (exact) approach to the exponential
summation involved in the cavity iteration and merging operations.%
\footnote{We note that a rearrangement of the sum over configurations
  according to the number of kinks looks like a natural perturbative
  expansion in $B_t$ for the system and naively scales polynomially
  $O(N_t^k)$ where $k$ is the maximum number of kinks
  summed. Unfortunately, little is gained from this rearrangement in
  terms of the accurately computable region of the phase diagram: the
  expected number of kinks for relevant rod configurations scales
  $\sim 2 B_t / \tau$ and thus the sum is again exponential in $B_t /
  \tau$.}  Fig. \ref{fig:phase-diag}(a) shows the phase diagram
calculated at $N_t = 10, N_{rods}=2500, N_{iter} \sim 1000 N_{rods}$
and suggestively fit to $N_t \rightarrow \infty$ using asymptotic
expansions in $1/N_t^2$. Qualitatively, all is as might be expected:
\begin{itemize}
\item At any $N_t$, the phase transition curve predicts a $B_t = 0$
  critical temperature in agreement with the analytic prediction of
  $\tau_c = 1/ \tanh^{-1}(1/\sqrt{q-1}) \approx 1.13$.
\item The upturn in the $N_t = 10$ phase boundary at low temperature
  is due to the finite discretization of time, which leads to an
  asymptotic phase transition line at $\tau = \frac{\tau_c}{N_t}
  \approx 0.113$.
\item While the fits to $N_t\rightarrow\infty$ are certainly
  approximate, we believe that the true $\tau = 0$ critical field lies
  between $1.5$ and $2$.  We believe continuous time techniques will
  allow dramatic refinement of this estimate and investigation of the
  quantum critical region.
\end{itemize}
Our phase diagram clearly disagrees with that of Kopec and
Usadel\cite{Kopec}, who treat the identical model using a soft
spherical approximation and find that both the critical temperature
and critical transverse field are depressed relative to our values.
Presumably this suppression of ordering arises due to the stronger
effect of fluctuations in the softened model.

Figure \ref{fig:phase-diag}(b) shows the instance averaged single site
von Neumann entropy $S_{vonN}$ which has a remarkably clear maximum
near the phase transition curve above the classical line. This
reflects the strength of quantum correlations even at the finite
temperature phase transition. See Ref.~\onlinecite{Arnesen} for
discussion of local measures of entanglement at finite temperature.

\begin{figure}[htbp]
\begin{center}
\includegraphics{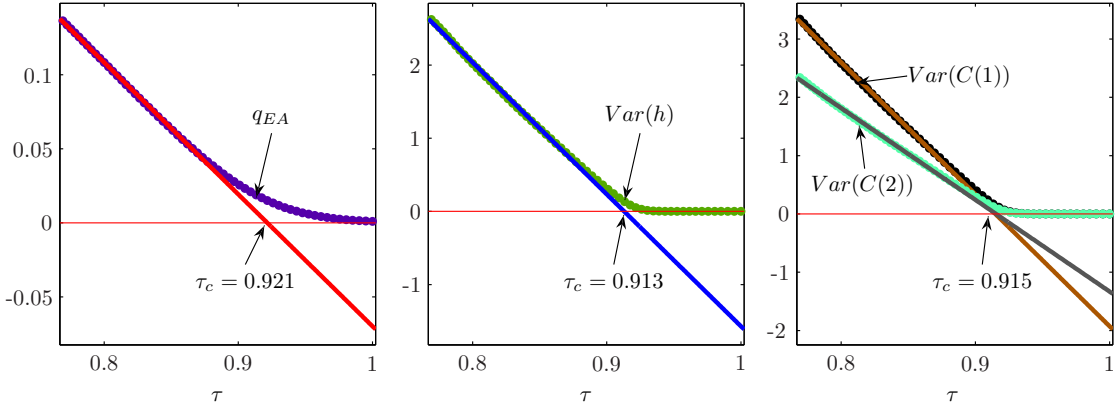}
\caption{These graphs all correspond to the horizontal slice at $B_t =
  1$ shown in Fig. \ref{fig:phase-diag}. The order parameter
  $q_{EA}$ and all effective action variances undergo mean field like
  transitions at the critical temperature (\emph{eg.} $q_{EA} \sim
  |\tau - \tau_c|^1$). This allows us to estimate the critical
  temperature precisely despite softening due to critical slowing down
  near the phase transition. These curves were calculated at $N_t =
  8$. However, the results are stable to increasing $N_t$ to $10$ to
  within an error of $\pm 0.005$ in $\tau_c$.}
\label{fig:pd10-varfit}
\end{center}
\end{figure}

Zooming in on the horizontal stripe at $B_t=1$ indicated on the phase
diagram, we find that $q_{EA}$ vanishes linearly at the critical
temperature (Fig. \ref{fig:pd10-varfit}(a)). This reflects the
underlying broadening transition in $P[\psi]$, which can be seen
sharply in the variances of each of the effective action coefficients
(Figs. \ref{fig:pd10-varfit}(b,c)). We use this behaviour to estimate
sharp transition points despite softening due to critical slowing in
the convergence of our procedure.

\begin{figure}[htbp]
\begin{center}
\epsfig{file=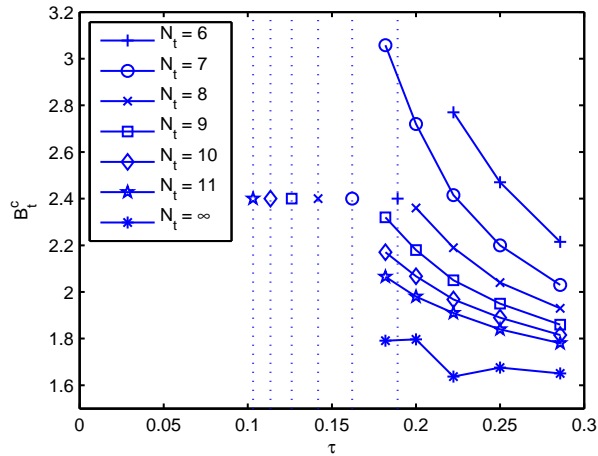,width=3.4in,height=2.55in}
\caption{The phase transition curve in the high field, low temperature
  regime at various $N_t$. The vertical dashed lines indicate the
  finite discretization critical asymptotes. The estimated curve for
  $N_t \to \infty$ in Fig. \ref{fig:phase-diag}a is given by fitting
  $B_t(N_t) = a / N_t^2 + b / N_t^4 + c$ to the points calculated at
  each temperature and then sending $N_t$ to infinity in the
  result. These fits suffer from a paucity of data, but are suggestive
  nonetheless.}
\label{fig:trans-curve}
\end{center}
\end{figure}

Finally, we note that much of the phase diagram is surprisingly stable
to variation in $N_t$. We have explored various regions of the phase
space at $N_t = 6,7,8,9,10,11$. The classical line ($B_t=0$) at all
temperatures is completely stable down to $N_t = 1$ as
expected. Perhaps more surprisingly, moving between $N_t = 8$ and $N_t
= 10$, $q_{EA}$ is essentially stable below $B_t = 1$ down to
temperatures $\tau \sim 0.15$. Of course, the high field, low
temperature part of the phase transition curve moves downward as the
finite discretization asymptote goes towards the $\tau=0$ axis. See
Fig. \ref{fig:trans-curve} for the low temperature critical curves
estimated using vertical stripes run at five different temperatures
(corresponding to $\beta = 3.5, 4, 4.5, 5, 5.5$) at various $N_t$.

\subsubsection{Structure of the Glassy Phase}

\begin{figure}[htbp]
\begin{center}
\epsfig{file=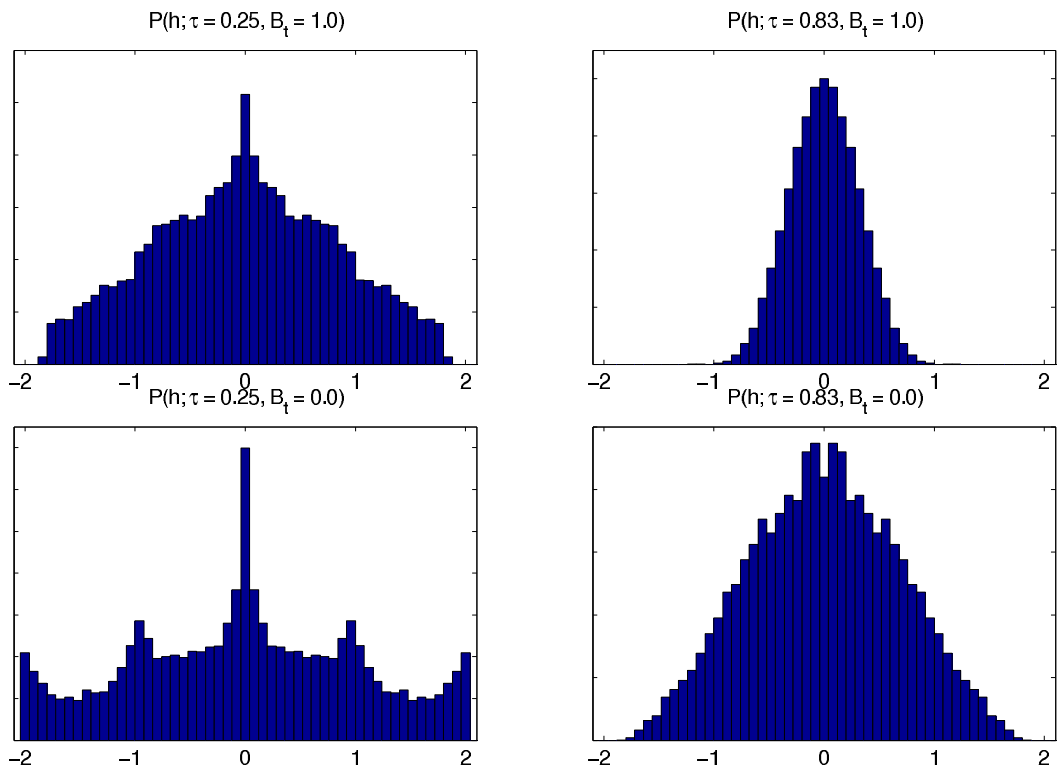,scale=0.75}
\caption{The distribution of the field term of the cavity rod action
  at the four different points in phase space labeled by the stars on
  Fig. \ref{fig:phase-diag}. The distinctive distribution of the low
  temperature replica symmetric classical spin glass is reproduced in
  the bottom left corner while the three other points all lie closer
  to the phase boundary in the $\tau$ or $B_t$ directions.}
\label{fig:four-points}
\end{center}
\end{figure}

Since the order parameter is a probability distribution of an action,
there is a rich structure to be investigated at even a single point
$(\tau, B_t)$ within the glassy phase. In Fig.
\ref{fig:four-points}, the marginal probability distribution of the
field term $h$ in the cavity action is shown at the four points
indicated on Fig. \ref{fig:phase-diag}(a). The two lower distributions
lie on the classical line ($B_t$ = 0), one deep within the glassy
phase and one near the transition. It is clear that the distinctive
features of the classical solution are reproduced here: a
Gaussian-like structure around $h=0$ near the phase transition with
the appearance of delta function spikes on the integer fields deep
within the phase. At $B_t = 1$, the qualitative picture of spread from
narrow Gaussian near the phase transition to broader, bumpier
distribution remains. It is less clear whether the sharply defined
spikes on integer fields would remain at $\tau = 0$ with large $B_t$.

\begin{figure}[htbp]
\begin{center}
\epsfig{file=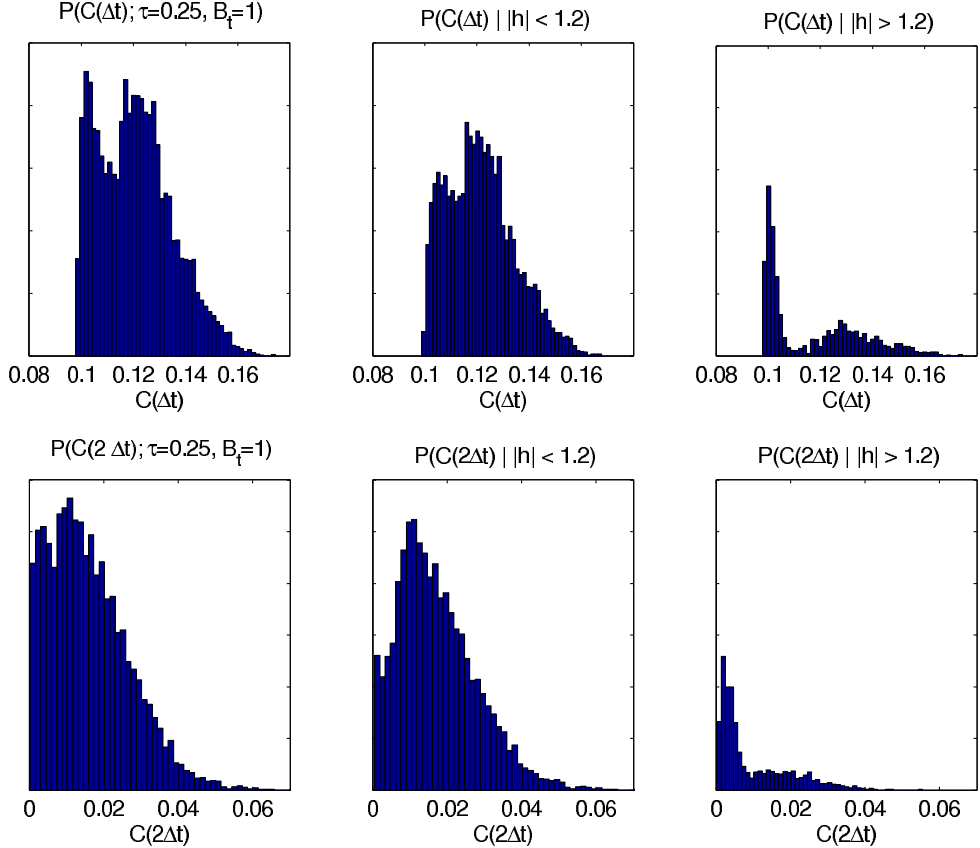,scale=0.8}
\caption{Histograms of nearest neighbor (top row) and next nearest
  neighbor (bottom row) interactions in imaginary time at $\tau =
  0.25$, $B_t =1$ (in the SG phase). The second and third column
  provide the conditional distribution of the interactions given $|h|$
  is small or large.}
\label{fig:hist-C12-b40-bt1}
\end{center}
\end{figure}

Further structure can be found in the nontrivial probability
distribution for the interaction terms that develop in the spin glass
phase. Figure \ref{fig:hist-C12-b40-bt1} shows the histogram for
various marginal and $h$-conditioned distributions of the nearest
neighbor and next nearest neighbor interactions terms at $(\tau =
0.25, B_t = 1)$ (cf. Fig.~\ref{fig:four-points}(top left)). We can
qualitatively understand many features of these distributions: 
\begin{itemize}
\item The two-spin interactions are ferromagnetic and the effect of
  coupling to neighboring rods is only to enhance the ferromagnetic
  interaction from the bare nearest neighbor interaction on a single
  rod ($\Gamma$). Indeed, this $\Gamma$ sets the minimum strength of
  $C(\delta T)$, as can be seen in the top row.
\item The strength of two-spin interactions are strongly
  anticorrelated with the strength of the cavity field $h$ as can be
  seen from the decomposition of the full marginal distributions of
  $C(\delta t)$ and $C(2\delta t)$ into the small (middle column) and
  large (right column) cavity field conditioned distributions. Large
  cavity fields on a central spin come from large fields biasing
  neighboring rods. These fields pin the neighboring spins more
  strongly and reduce the ability of those spins to mediate
  interactions in time between the central rod spins, reducing the
  effective two-spin interaction.

\item The multimodal spikiness in these distributions reflect
  the spikiness in the low temperature cavity field distributions
  through the field-interaction correlation.
\end{itemize}
Unfortunately, we have yet to develop a more significant analytic
understanding of these correlations nor a means to extrapolate them to
zero temperature in the presence of the transverse field. 

\begin{figure}[htbp]
\begin{center}
\epsfig{file=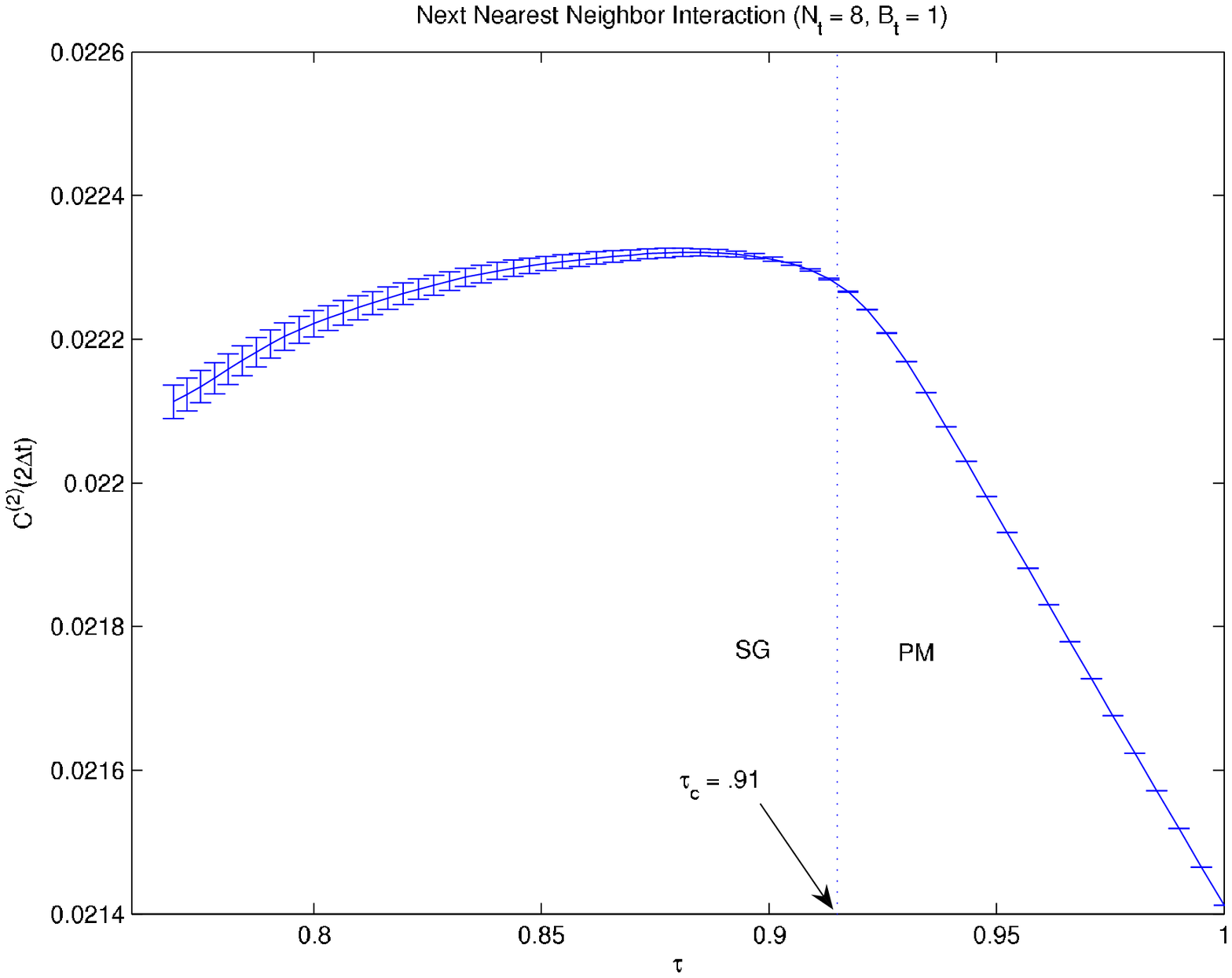,scale=0.45}
\epsfig{file=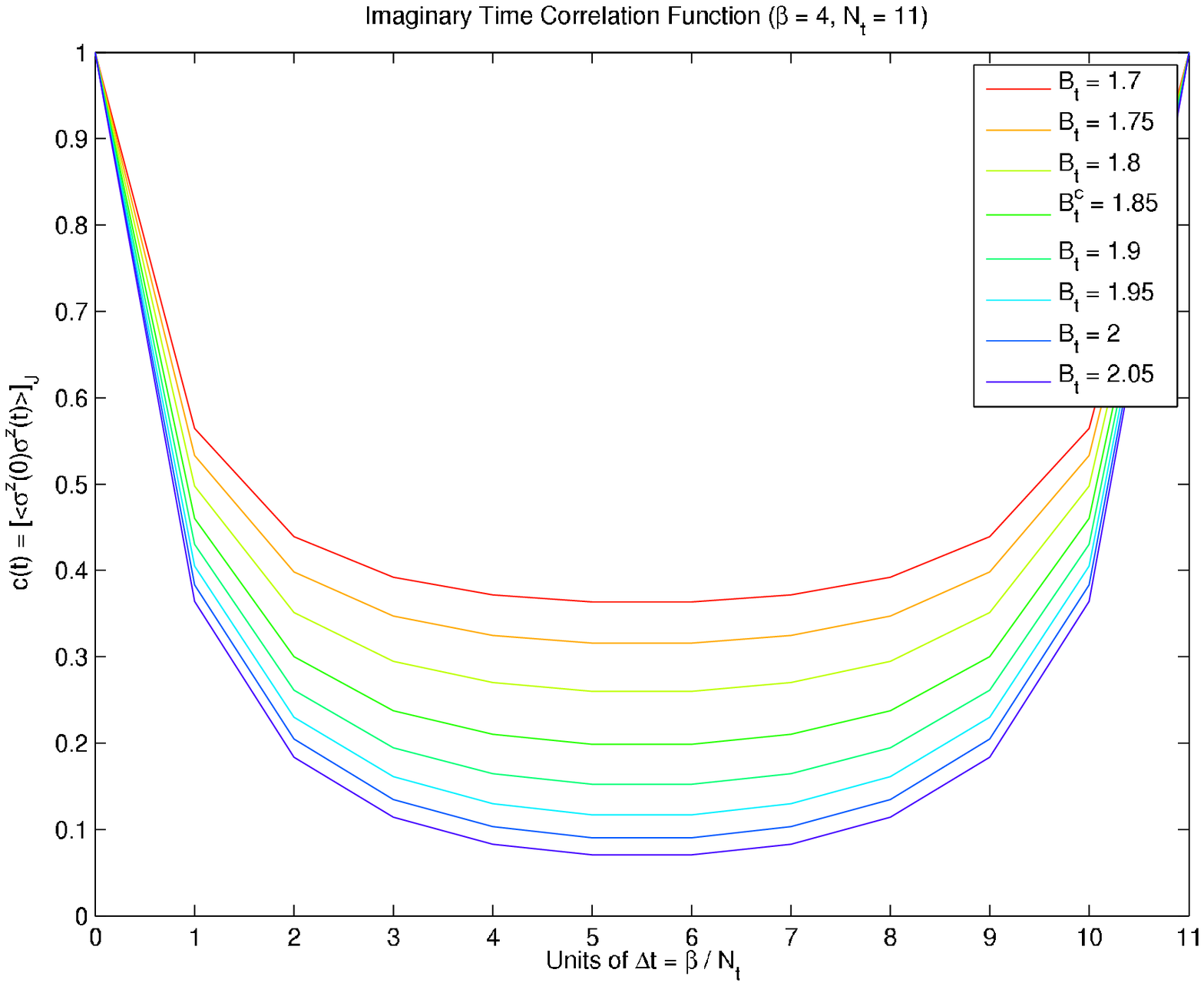,scale=0.45}
\caption{(a) The average next to nearest neighbor interaction in time
  $C^{(2)}(2\Delta{}t)$ of a cavity rod action at $B_t = 1$, varying
  the temperature. The bars indicate the variance of
  $C^{(2)}(2\Delta{}t)$. Notice the zero variance above the critical
  temperature. (b) Single site imaginary time correlation function
  $\avg{\sigma(0) \sigma(t)}_J$ at $\tau = 0.25$ for various $B_t$
  passing through the transition at $B_t^c \sim 1.85$.}
\label{fig:c-imag-time}
\label{fig:C2-vs-tmp}
\end{center}
\end{figure}

Finally, we emphasize that the phase transition is signaled by a
singular broadening of $P$ rather than any singularity in its first
moments or in the structure of the typical imaginary time action.
This can be seen in the smooth evolution of $\avg{C(2\Delta{}t)}_J$
through the phase transition in Fig. \ref{fig:C2-vs-tmp}a and the
similarly smooth evolution of the average two point correlation
function $\avg{\sigma(0) \sigma(t)}_J$. This is in contrast to the
ferromagnetic case in which the distribution $P(\psi)$ would exhibit
spontaneous symmetry breaking but remain deterministic.

\section{Discussion and further work}

The cavity method has a long and illustrious history in the study of
statistical systems -- from Bethe's early work on the Ising
ferromagnet to modern studies of random constraint satisfaction
problems in computer science. In this paper, we have introduced a new
variant for studying disordered quantum systems within an imaginary
time formalism. This represents an intuitively appealing, natural
synthesis of the classical disordered model techniques  with
the quantum homogeneous models studied in DMFT. We note that our
framework can be simply adapted to study many other transverse field
Ising systems on trees with fixed or fluctuating connectivity -- such
as the ferromagnet, diluted ferromagnet or biased glass.

We have shown that the transverse field Ising glass on a Bethe lattice
of connectivity three has a phase transition line all the way from the
classical $B_t=0,T=1.13$ to the quantum $B_t\approx 1.75,T=0$. At
finite temperature, the transition is classical and mean field
like. Inside the frozen phase the picture is similar to the classical
case: when a randomly chosen spin is extracted from the graph, its
effective action (analogous to the cavity field) is well-defined in
the paramagnetic region but is a random functional in the spin-glass
phase. In this phase all of the local observables, classical and
quantum, are therefore also random variables. In principle one could
also study the properties of the entanglement of distant spins, which
cannot be done in the better known but fully connected SK model.

The reader familiar with classical spin glass theory will have noticed
our avoidance of the important issue of replica symmetry breaking
(RSB), which is widely believed to be a feature of a correct treatment
of the random graph Bethe lattice. In this connection, we note that
our simpler treatment is indeed correct for models on Cayley trees
with fixed boundary conditions. Quantum RSB phenomena certainly
deserve further study: the conceptual difficulties of the RSB ansatz
become even thornier in the presence of quantum tunneling. We note
that formally breaking replica symmetry at the one step level (1RSB)
should be straightforward in our framework. In analogy with
Ref.~\onlinecite{MP}, one should introduce \emph{populations} of
populations of effective actions and weigh them according to their
free energy (as one does for different solutions of the TAP
equations). This straightforward modification of the algorithm makes
it computationally considerably more time consuming. For this first
pass, we decided not to embark on such a project.

Unfortunately, we are not aware of Quantum Monte Carlo or other
numerical studies on the transverse Ising model on the Bethe lattice
with which to compare our results. The primary difficulty that has
prevented the direct simulation of this system is that it requires a
very large number of spins to approximate the infinite system
effectively. Since a random graph with fixed connectivity has an
extensive number of loops of lengths $\ell\sim{\cal{O}}(\ln N/\ln
(q-1))$, $N$ needs to be exponentially larger than any statistically
relevant length scale (\emph{e.g.} the coherence length).
  
Another direction for further development would be to find a spin
glass (or otherwise) model amenable to analytic treatment within the
quantum cavity method. A ``soft spin'' Gaussian model would do, since
the path integrals to be performed in an iteration could be computed
exactly. However, we do not believe this model has a spin glass phase
in the absence of higher order couplings. Whether one could treat such
a coupling perturbatively is a question worthy of further exploration.

\section{Acknowledgments}

We would like to thank M.\ Aizenmann, R.\ Bhatt, D.\ Huse for
discussions. C.L.\ and A.S.\ would like to thank G.\ Santoro, R.\
Fazio, E.\ Tosatti and R.\ Zecchina for their kind hospitality at ICTP
and SISSA in Trieste, where part of this work was completed, and for
discussions. A.S.\ would like to thank E.\ Farhi and J.\ Goldstone for
discussions during the early stages of this project.

Note added: During the production of this manuscript, several other
groups\cite{Hastings,PoulinLeifer} have independently developed
quantum cavity techniques inspired by somewhat different approaches to
the quantum problem. Ref. \onlinecite{PoulinBilgin} provides a
quantitative comparison of the error in several of these related
techniques. In an important development, Ref. \onlinecite{Krzakala}
describes a continuous time version of the quantum cavity method
described here which they apply to a ferromagnetic model.

\end{document}